\documentstyle[12pt,aaspp4]{article}

\def\clock{\count0=\time \divide\count0 by 60
     \count1=\count0 \multiply\count1 by -60 \advance\count1 by \time
     \number\count0:\ifnum\count1<10{0\number\count1}\else\number\count1\fi}

\begin{document}

\title{Generating Cosmological Gaussian Random Fields}
\author{Ue-Li Pen}
\affil{Harvard-Smithsonian Center
for Astrophysics, 60 Garden St., Cambridge, MA 02138}
\newcommand{\etal}{{\it et al.}}

\begin{abstract}
We present a generic algorithm for generating Gaussian random initial
conditions for cosmological simulations on periodic rectangular
lattices.  We show that imposing periodic boundary conditions on the
real-space correlator and choosing initial conditions by convolving a
white noise random field results in a significantly smaller error than
the traditional procedure of using the power spectrum.  This
convolution picture produces exact correlation functions out to
separations of L/2, where L is the box size, which is the maximum
theoretically allowed.  This method also produces tophat sphere
fluctuations which are exact at radii $ R \le L/4 $.  It is equivalent
to windowing the power spectrum with the simulation volume before
discretizing, thus bypassing sparse sampling problems.  The mean
density perturbation in the volume is no longer constrained to be
zero, allowing one to assemble a large simulation using a series of
smaller ones.  This is especially important for simulations of
Lyman-$\alpha$ systems where small boxes with steep power spectra are
routinely used.

We also present an extension of this procedure which generates exact
initial conditions for hierarchical grids at negligible cost.
\end{abstract}


\section{Introduction}

Cosmological N-body and hydrodynamic simulations have numerical
resolution limitations imposed on both large scales comparable to the
simulation box (Gelb and Bertschinger 1994, hereafter GB), as well as
on small scales due to finite resolution (Hockney and Eastwood 1981).
For simulations of Cold Dark Matter initial conditions, they found
that for a box of width L=51.2 Mpc, the truncation of the power
spectrum at the box length resulted in fluctuations on 8 Mpc spheres
$\sigma_8$ which are systematically 40\% lower than the analytical
value.  This factor of nearly 2 in error arises due to the truncation
of the power spectrum while generating the initial conditions.  They
found that a box size with L=100 Mpc was required to reduce the error
to 10\%.  In cosmological applications, the actual value of $\sigma_8$
is of significant interest since it corresponds to the mass
fluctuations from which a rich cluster forms.  In the Press-Schechter
framework (Press and Schechter 1974), cluster abundances will be
statistically correct as long as $\sigma_R$ is exactly modeled in a
simulation, where $M=(4/3)\pi R^3 \rho_c$ is the mass of the cluster
in question.  Since computational cost grows at least as fast as the
third power of the simulation volume at fixed physical resolution, it
is desirable to use numerical discretizations which allow the most
accurate representation of physical processes with the minimal
simulation volume.  The first purpose of this letter is to demonstrate
how the initial conditions in a periodic box can be chosen to
eliminate this error entirely for tophat smoothing scales which are of
size up to L/4.  When an ensemble of simulations is performed, their
average will reflect the correct tophat fluctuation distribution.

The current trend in high-resolution simulations is to adaptively place
high resolution grids in regions which are known to form high density
objects (Bryan and Norman 1995, Dutta 1995, Navarro
\etal\ 1995, Bartelmann and Steinmetz 1996).  If this procedure is
repeated recursively, one needs to be able to specify the initial
conditions at very high resolution in these regions.  The brute force
approach needs to generate the initial conditions at high resolution
everywhere, and throw out the regions which are never resolved.  The
convolution picture (Salmon 1996) allows a simple decomposition of the
problem into multiple scales, each of which is rapidly computed using a
FFT.  This paper describes the optimal implementation of the
convolution picture for cubical regular or hierarchical grids.

A third related  problem is the question of how one might embed a
small periodic simulation volume inside of a larger one.  For example,
one might simulate a large $1 $Gpc$^3$ volume of the universe at
modest resolution, and identify a region in this simulation (perhaps
an extremely rich cluster or empty void) which one would like to study
at higher resolution.  Since perturbations on scales of even 100 Mpc
are small, one might envision drawing 100 Mpc box from the large
volume and only simulating the smaller volume at higher resolution.
If the simulation imposes periodic boundary conditions, the procedure
described in this letter will allow us to generate such a box which
agrees very well with the larger box in the central regions of
overlap.  The converse was studied by Tolman and Bertschinger (1996),
which is how to run a large simulation at high resolution given a
small one.  This problem also fits in the framework of the convolution
picture.

The general solution to all three problems is to envision the
generation of initial conditions in coordinate space instead of
Fourier Space.  We will initially lay down a random number at each
grid point.  Now gridpoints are uncorrelated with each other, and the
correlator is a Dirac $\delta$-function.  Let us now convolve this
initial condition with some appropriate window $W_\rho(r)$.
Convolution, being a linear process, will preserve the Gaussianity of
the random field, but introduces a correlation function which is just
the auto-correlator of $W_\rho$.  Numerically, this convolution will be
performed again with the aid of FFTs.  Since it is so
similar to the Fourier method in implementation one can ask why it
would give significantly 
different answers.  The standard discretization of the power spectrum
only samples at discrete frequencies $k$.  For small $k$, this error
can be large, especially if the slope of the power spectrum is steep.
Discretizing a periodic correlator is equivalent to convolving the
power spectrum with the window function of the simulation geometry,
which smoothly averages over the power spectrum in the regions were it
is poorly represented.  For power spectra which have significant
features, for example due to baryon oscillations, this is especially
important.  The correlator constructed in this manner should show
absolutely no error for $r<L/2$.  This is the qualitative reason for
the improved performance of the convolution picture.

The effect of truncating the power spectrum are particularly severe if
the power spectrum is steep.  As the slope $n$ of the power spectrum
$P(k)=k^n$ approaches $n \rightarrow -3$, most of the small scale top
hat variance arises from large wavelength modes which are not included
in the box size.  For Cold Dark Matter power spectra, the slope
approaches this value on small scales, and simulations of the early
non-linear mass scales, for example in Lyman-$\alpha$ simulations, are
more strongly affected by the loss of long waves (see for example
Rauch \etal\ 1996 and references therein).

This letter will mathematically describe the implementation in section
\ref{sec:math}.  Numerical measures of the convolution performance are
given in section \ref{sec:num} together with a discussion of its
application to generate simulations which give the correct ensemble
average.  In section \ref{sec:subgrid} we describe how the convolution
method allows the seamless adaptive generation of subpower for
hierarchical grids, and how to generate consistent periodic
sub-(super) boxes.  Initial conditions on unstructured grids can be
treated by the tree algorithm (Salmon 1996).

\section{Formalism}
\label{sec:math}
\newcommand{\bx}{{\bf x}}
\newcommand{\bk}{{\bf k}}

We consider density fields $\delta(\bx)$ where $\bx=(x,y,z)$ is a
position in three dimensional Euclidean space.  The {\it
autocorrelation function} (or auto-correlator for short) is defined as
$\xi(r)=<\delta(\bx) \delta(\bx+r)>$.  The density field can be
transformed into Fourier space $\delta(\bk) = V^{-1}\int \exp(-i\bk \cdot
\bx) \delta(\bx) d^3x$, where we measure the {\it power spectrum}
$P(k)=\delta(\bk)\delta(-\bk)$.  The power spectrum is also the Fourier
Transform of the auto-correlator.  We define the spherical tophat window
function $W_T(u)=3[\sin(u)-u\cos(u)]/u$ with which we measure the
standard deviation of the density field smoothed on tophat spheres of
radius $R$ 
$
\sigma_R^2 = 4\pi\int P(k) W_T(k R) k^2 dk.
$
The {\it
traditional discretization} to obtain a random density field is to
choose a Gaussian random number for each mode $\delta(\bk)$ in Fourier
space with variance $P(|\bk|)$.  For small $k$, the spacing of modes
$\Delta k \sim 2\pi/L$, where $L$ is the simulation box length.  Where
$k \sim 2\pi/L$, this sampling introduces significant coarse graining
error, which we will call the {\it truncation error}.  Within the
Fourier domain, we next introduce the cubical tophat window $W_C(\bk) =
\sin(2\pi L k_x)\sin(2\pi Lk_y)\sin(2\pi Lk_z)/[(2\pi L)^3 k_xk_yk_z]$,
which reflects the simulation geometry.
To implement the {\it convolution picture}, we define the {\it
correlation kernel} $W_\rho(k) = \sqrt{P(k)}$.  We call it a
``picture'' since it is equivalent to the Fourier space discretization
for infinite volumes, so no new physics is introduced.  Since $P(k)$
is positive, so is $W_\rho$, but not necessarily its inverse Fourier
transform $W_\rho(\bx)$.  One can verify that the auto-correlator is
the auto-convolution of the correlation kernel $\xi(r) = <W_\rho(\bx)
W_\rho(\bx+r)>$.  We now generate a random field with zero correlation
length $\delta_0(\bx)$ and define $\delta(\bx) = \int \delta_0(\bx')
W_\rho(\bx-\bx') d^3\bx'$.

We now consider two alternatives to implement these relations on a
periodic lattice.  In Fourier space, periodicity means that only a
discrete spectrum of modes are represented.  At the modes which are
represented, the traditional approach is to simply use the power
spectrum $P(k)$ at those discrete intervals.  Two problems arise.  If
the power spectrum changes significantly between successive modes, one
misses the intermediate power.  The extreme case would be if the power
spectrum had a feature, say a peak, which did not lie on one of the
discrete grid points.  The other problem is that the $k=0$ mode is now
forced to be zero.  We see that we really want some way of not just
using the value of $P(k)$, but some average over the width of the
simulation volume $W_C$.

In the convolution picture, we impose periodicity on the correlator.
We first set the correlator equal to zero outside of the box volume,
and then create periodic images.  We will add a superscript $^{\rm
grid}$ to indicate any quantity which has been made periodic using
this convolution picture prescription.  For the effective grid power
spectrum, this is equivalent to replacing the simulation power by the
full power spectrum convolved with $W_C$:
\begin{equation}
P^{\rm grid}(\bk) = \int P(|\bk'|) W_C(\bk'-\bk) d^3\bk'.
\label{eqn:pc}
\end{equation}
The effect of this convolution is biggest at small $k$, especially
$k=0$, where the analytic power spectrum is typically 0, but the
windowed one takes on some finite value which is similar to the
variance in the box simulation volume.  The window function of the
simulation geometry smoothly averages between the finite discrete
sampling imposed by the periodic boundary conditions.  
Due to the anisotropy of $W_C$  the grid power
spectrum $P^{\rm grid}(\bk)$ has an angular dependence.  We will
discuss the effects of (\ref{eqn:pc}) in more detail in section
\ref{sec:num}.

\section{Errors}
\label{sec:num}

\begin{figure}
\plotone{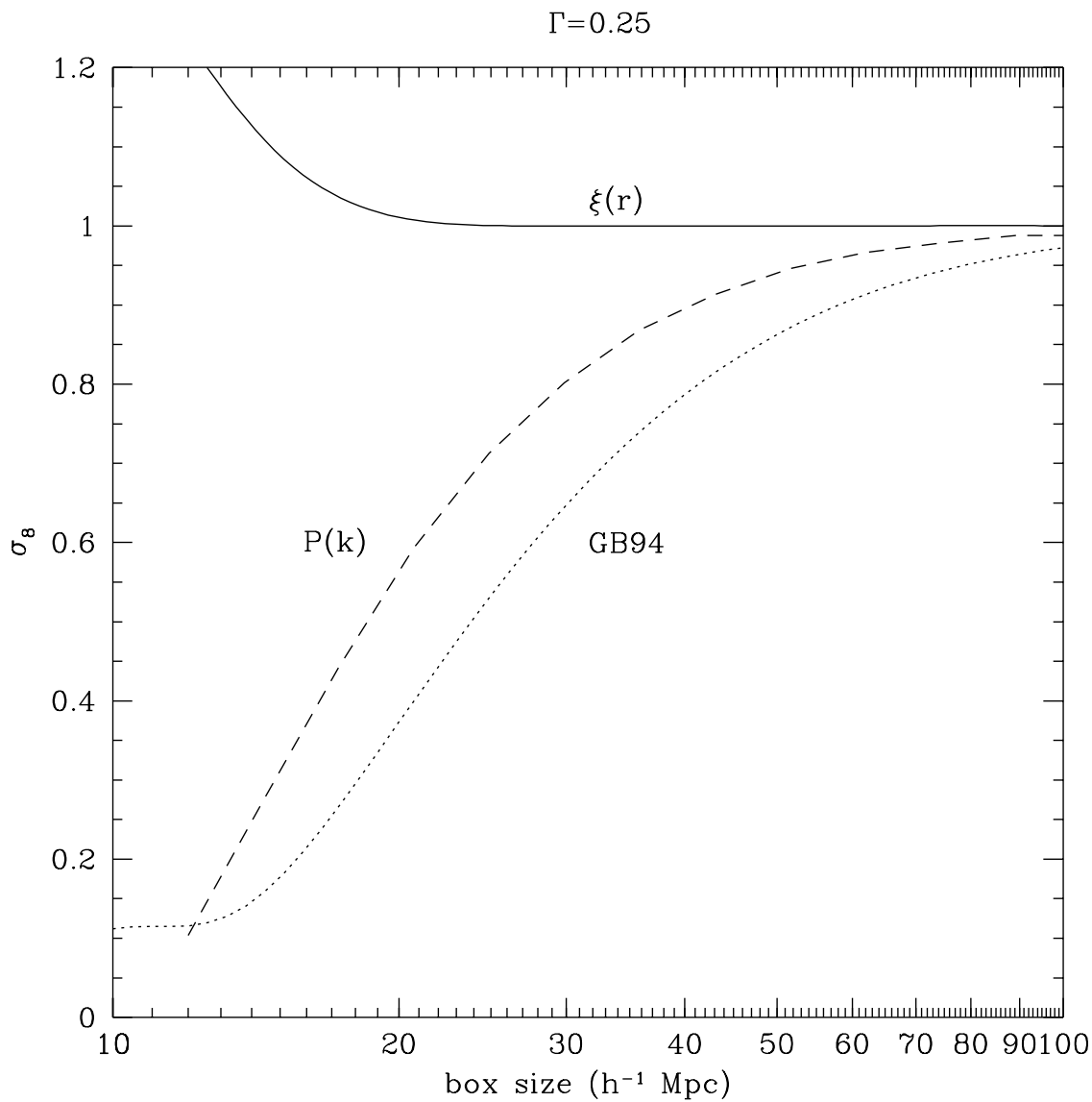}
\caption{\protect{$\sigma_8$} as a function of box size using various
approximations.  The dotted line was computed by GB94 using a truncated
power spectrum.  The dashed line is the actual tophat fluctuation in
a $128^3$ box.  The solid line was obtained using the convolution
picture.  The latter is exact for box sizes greater than 32 $h^{-1}$ Mpc.}
\label{fig:sigma8}
\end{figure}

Figure \ref{fig:sigma8} shows the ensemble averaged value of
$\sigma_8$ computed using the traditional power spectrum as well as in
the convolution picture.  The GB estimate of $\sigma_8$ truncates the
lower bound of the integral over the power spectrum in Equation
(\ref{eqn:s8}) at $k=2\pi/L$.  The result for a model with power
spectrum $\Gamma=.25$, $\sigma_8=1$ (Bardeen \etal\ 1998) is shown as
the dotted line in the figure.  In order to obtain the exact value of
the periodic box $\sigma_8$, we computed the actual tophat convolution
on a discrete $128^3$ grid without assuming a spherical cutoff in the
power spectrum.  This leads to the actual values of $\sigma_8$ shown
as the dashed line.  The solid line from the convolution picture was
generated on a $256^3$ lattice taken from a code which actually
computes N-body simulation initial conditions.

We notice the significant improvement in the linear theory analysis
which is exact for box sizes greater than 32 $h^{-1}$ Mpc.  Even for
box sizes of 16 $h^{-1}$ Mpc, where the 8 $h^{-1}$ Mpc spheres start
wrapping around the period volume, the convolution picture has an
error of only 7\%.  By contrast, discretizing the power spectrum
introduces an error of over 60\%.  We notice further that the
convolution picture typically {\it overproduces} $\sigma_8$, while
truncation in Fourier space underestimates it.  We can understand it
as follows.  The correlation function is now forced to be periodic, so
instead of steadily decaying at large separations, it increases beyond
the periodicity radius.  We have
\begin{equation}
\sigma_R^2 = 4\pi\int_0^{2R} \xi(r) W^2_T(r) r^2 dr
\label{eqn:s8}
\end{equation} 
where $W^2_T(r)$ is the convolution of the tophat kernel with itself.
It is zero for $r>2R$, so we say that it has {\it compact support}.
Periodicity thus causes (\ref{eqn:s8}) to be overestimated because
$\xi$ itself is overestimated for correlators which are positive and
decreasing at $r=L/2$.  Truncation in Fourier space, on the
other hand, usually results in an underestimate because long waves are
lost.

\begin{figure}
\plotone{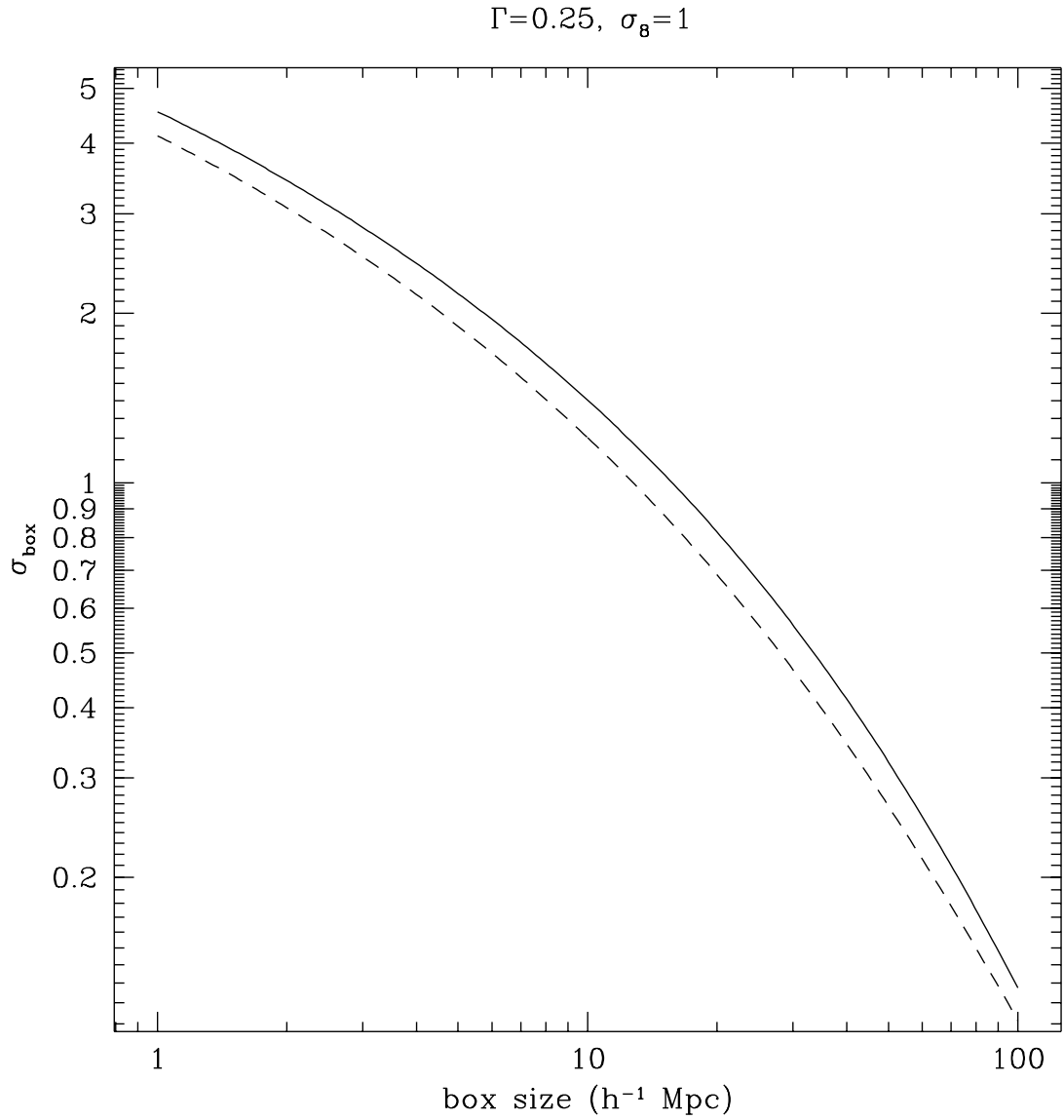}
\caption{The standard deviation of the box density using the
correlator (solid) and the exact value (dashed).  The normalization
chosen is $\sigma_8=1$ and $\Gamma=0.25$.  The convolution picture
always slightly overestimates the box-to-box variance (see text).
}
\label{fig:boxvar}
\end{figure}

A related problem is the box-to-box density variance.  In the
traditional power spectrum generation, the total density within each
box is typically defined to be the cosmic mean density.  In reality,
this value varies even when averaged over a corresponding volume of
the universe.  The mean density of the box can be modeled as a change
of the mean matter density $\Omega$ and incorporated in the evolution
of the scale factor.  This addresses one of the causes of
systematically underestimating $\sigma_8$, which can result in
systematic errors when simulating the abundance of virialized objects
such as clusters of galaxies in finite volumes.  In the convolution
picture, the periodic correlator does not have zero mean, so the $k=0$
mode will have finite power.  The variance thus induced
will not be exactly equal to the true variance in cubical tophats of the same
volume.  The difference in shown in Figure \ref{fig:boxvar}.  We see
that the variance of the whole box variance is slightly overestimated, but
typically only by 10\%.  The tophat spheres of
up to $R=L/4$ all have the exact variances, but the box variance
itself corresponds to $R\gtrsim L/2$.  The correlator is already
completely determined by requiring the smaller ones to be exact, so we
have no additional freedom to adjust the large scale fluctuations.
Luckily the corrections are small because the dominant contribution in
(\ref{eqn:s8}) comes from $R<L/4$.

A dual view of this picture is that the box-to-box variance is
described by Equation (\ref{eqn:pc}) where $W_C$ is replaced by
$W_C^2$.  One could pose the question why one doesn't simply make such
a replacement for all modes.  The reason is that imposing exact
variances in tophat spheres which can be embedded in half the
simulation width, makes Equation (\ref{eqn:pc}) the unique solution.
If one is interested in the dynamics and abundance of bound objects,
the success of the Press-Schechter ansatz argues that the top-hat
spheres should be of primary importance.  A potential drawback of
Equation (\ref{eqn:pc}) is that the grid power may become negative,
making it impossible to use it as a variance.  On the numerical grid,
one can add a constant to the numerical power spectrum to make it
positive everywhere.  This is equivalent to adding white noise, which
shows up only in the $r=0$ bin of the two-point correlator.  In
practice, for the transfer function used in this study, the effect of
varying $\xi(0)$ between 0 and $\xi(\Delta r)$, the value of the
correlator at one grid cell lag, did not affect measures like
$\sigma_8$ by more than 1 part in 1000 even down to box sizes of 10
Mpc on a $128^3$ grid.  We finally note that the simulation results
must be analyzed in the same way as an observational survey.  The
power spectrum of the simulation volume is not a true representation
of the infinite volume power spectrum.  It is instead a windowed
sample, which must be taken into account during statistical analysis.

\section{Subgrid Power}
\label{sec:subgrid}

\begin{figure}
\plotone{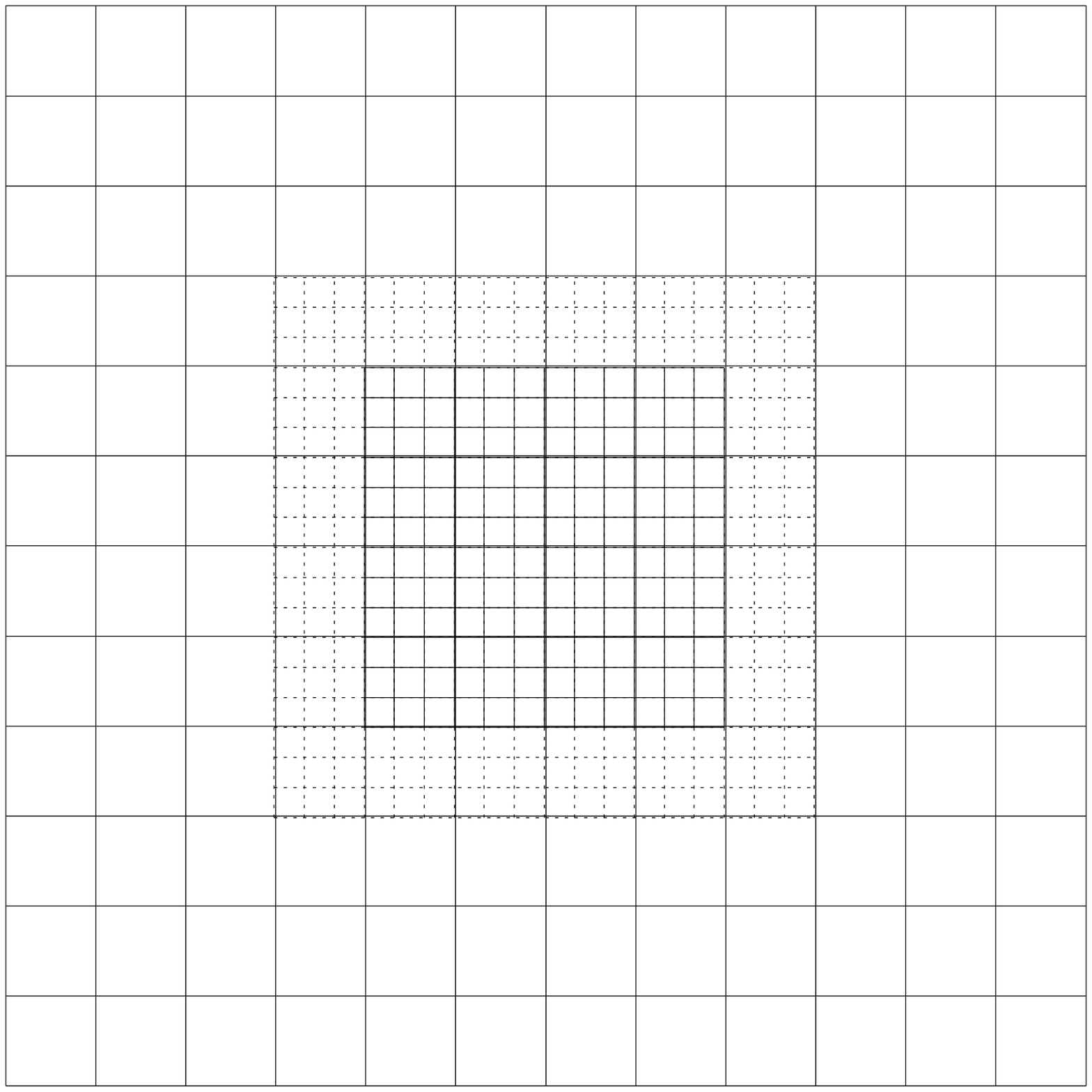}
\caption{Hierarchical mesh.  The dotted lines are the buffer zones on
which fine mesh initial seeds need to be chosen such that a seamless
period-free convolution can be applied.
}
\label{fig:subgrid}
\end{figure}

One direct advantage of the convolution picture is its adaptability
to arbitrary domains.  We first consider the situation where we want
to add power on subgrid scales.  Consider a fine rectangular grid
(heavy lines) which is superposed on a coarser outer grid as shown in
figure \ref{fig:subgrid}.  Also shown is the dotted extended fine
grid, which is larger than the fine grid by an amount parameterized by
$R$.  We will assume that a random initial condition was already
generated on the coarse grid.

We first note that the correlation kernel can always be decomposed
into a sum of two kernels
$
W_\rho=W_R+W_R'
$
where we can choose 
\begin{equation}
W_R(r) = \left\{ \begin{array}{ll} W_\rho(r) \ \ \ \ &{\rm if \ \ \ \ r>R} \\
			    W_\rho(R) & {\rm if \ \ \ \ r \le R} 
		 \end{array}	  
	 \right.
\end{equation}
By deduction, $W_R'$ will be zero for $r>R$, i.e. it has compact
support.  The correlation kernels are obtained by taking the square
root of the periodic correlator $W_{\rho,\rm grid}(k) = \sqrt{\xi_{\rm
grid}(k)}$ in Fourier space and transforming back to coordinate
space.  

We will first generate initial conditions on the whole coarse grid
with $W_R$.  We linearly project these values onto the fine grid.
This we will call $\delta^{\rm fine}_1(\bx)$.  We then Fourier
Transform all the white noise conditions $\delta_0(\bk)$ into
coordinate space on the coarse grid $\delta_0(\bx)$, and also project
these seeds onto the extended fine grid which we call $\delta_0^{\rm
fine}(\bx)$.  Let N be the refinement factor.  The fine grid is now
constant on cubicles with $N^3$ elements.  For each cubicle, we will
add random white noise on each grid cell of variance $N^3$ times the
original white noise variance, subject to the constraint that the
average must be the coarse cell value.  Call the white noise field
$\delta^{\rm fine}_2$.  To implement the constraint, we subtract from
each cubicle its mean and add the coarse cell value.  Call the
resultant density field $\tilde\delta^{\rm fine}_2$.  We now convolve this
field with $W_R'$ and add it to $\delta^{\rm fine}_1$.  Since $W'_R$
is only nonzero for $r<R$, we only need to add a buffer zone of width
$R$ to the fine grid to obtain a full fine grid power spectrum.  The
actual choice of $R$ depends on how smoothly one wishes to interpolate
from coarse grid power to fine grid power.  A typical choice of 2 grid
cells should interpolate reasonably smoothly.


\newcommand{\cp}{{\cal P}}
\newcommand{\cR}{{\cal R}}

In the language of multigrid (Press \etal\ 1992), we can summarize this
procedure in terms of the prolongation operator $\cp$, the restriction
operator $\cR$ and the convolution operator $\ast$.  $\cp$ maps from
the coarse grid to the fine grid by using the nearest grid point.
${\cal R}$ does the converse map by averaging over the contained
subdomain.  We define the constrained fine grid as a white noise
random field with variance $<(\delta^{\rm fine}_2)^2> = N^3
<\delta_0^2>$, and $\delta_0={\cal R} \delta^{\rm fine}_1$.  Using the
new operators, we can implement the constrained subpower as
$\tilde\delta^{\rm fine}_2=\cp\delta_0+\delta^{\rm fine}_2 - \cp \cR
\delta^{\rm fine}_2$. This gives us a compact prescription for the new
fine grid power:
\begin{equation}
\delta^{\rm fine} = \cp W_R \ast \delta_0 + W_R' \ast
(\cp\delta_0+\delta^{\rm fine}_2 - \cp \cR \delta^{\rm
fine}_2 ).
\end{equation}

The same procedure can be used to embed a high resolution small volume
simulation inside a larger grid.  The reverse of the procedure above
will be used.  We build the large grid white noise field from the
small grid using the restriction operator $\delta_0 = {\cal R}
\delta_0^{\rm fine}$ in the region of overlap, and fill the rest of
the volume with new random numbers.  These are then convolved with the
appropriate window $W_\rho$.  If the power on the grid scale of the
small volume simulation was moderately small, the two grids will be
consistent with each other in the central regions of overlap.  The
converse operation corresponds to cutting out a small volume from a
larger one, and simulating the small volume at high resolution with
periodic boundary conditions.  This is straightforward if we perform
the cutting on the white noise grid $\delta_0(\bx)$, and then apply
the periodic convolution on the smaller white noise grid.  

\section{Conclusion}

Numerical simulations introduce unphysical truncation errors on two
scales: the grid scale and the box length.  At the grid scale, errors
are dominated by the computation of forces and field evolution.  At
the box scale, currently simulations are often limited by the
discretization of the power spectrum.  For many problems in cosmology,
this limit can be improved when described in the picture of
Press-Schechter structure formation.

We have shown how initial conditions should be generated for
cosmological simulations utilizing small periodic box volumes.
A box is considered small if the variance on the box scale is not
small, typically greater than $10\%$.  For CDM models, this means any
box smaller than $\sim 50 h^{-1}$ Mpc.  By discretizing the correlator
instead of the power spectrum, we obtain more accurate estimates of
tophat variances and correlators, which are often the primary target
of study in N-body simulations.  

We have also applied this approach to decompose the correlation kernel
into a long range and short range part, allowing us to add small scale
power for hierarchical subgrids without imposing periodicity on the
subgrid subpower.  This procedure generalizes to embedding small
periodic simulations inside of larger ones and vice versa.

I would like to thank Ben Bromley, Bill Press and Uros Seljak for
useful discussions.  This work was supported by the Harvard Society of
Fellows.


\begin{thebibliography}{}
\bibitem{}  Bardeen, J.M., Bond, J.R., Kaiser, N., \& Szalay, A.S. 1986
\apj, 304, 15
\bibitem{} Bartelmann, M., Steinmetz, M. 1996, \mnras, 283, 431.
\bibitem{} Bryan and Norman 1995, BAAS, 187, 95.04.
\bibitem{} Dutta, S.N. 1995, \mnras, 276, 1109.
\bibitem{} Gelb and Bertschinger 1994, \apj, 436, 491.
\bibitem{} Hockney, R.W. and Eastwood, J.W. 1981, ``Computer
Simulations using Particles'', New York: McGraw Hill.
\bibitem{} Miralda-Escude, J., Cen, R., Ostriker, J. P., Rauch,
M. 1996, \apj, 471, 582.
\bibitem{} Navarro, J.F., Frenk, C.S., White, S.D.M. 1995, \mnras,
275, 56.
\bibitem{} Pen, U. 1996, astro-ph 9610147, submitted to \apj.
\bibitem{} Press, W.H. and Schechter, P. 1974, \apj, 187, 425.
\bibitem{} Press, W.H, Teukolsky, S.A., Vetterling, W.T., Flannery, B.P.,
1992, Numerical Recipes (Second Edition). Cambridge
University Press, Cambridge
\bibitem{} Salmon, J. 1996, \apj, 460, 59.
\bibitem{} Tolman and Bertschinger 1996, \apj, 436, 491.

\end{thebibliography}
\end{document}